# Full vectorial field sensing using liquid crystal droplet arrays

Xuke Qiu[1,†], Jinge Guo[1,†], Jiahe Cui[1], Runchen Zhang[1], Zimo Zhao[1], Yifei Ma[1], Waqas Kamal[1], Steve J. Elston[1], Alfonso A. Castrejón-pita[1], Stephen M. Morris[1,*], and Chao He[1,*]

[1]Department of Engineering Science, University of Oxford, Parks Road, Oxford, OX1 3PJ, UK

[†]*These authors contributed equally to this work*
[*]*Corresponding authors:* stephen.morris@eng.ox.ac.uk; chao.he@eng.ox.ac.uk

## Abstract

**Determining the amplitude, phase, and polarization profile of light is essential for both fundamental scientific discovery and applications spanning optical metrology, microscopy, astronomy, and optical communication/computing technologies. However, most modern measurement approaches are unable to retrieve such parameters readily, often relying on bulky and expensive hardware, or lacking the capability for single-shot sensing. Here, we introduce a low cost, compact, full vectorial field sensor based on an inkjet-printed nematic liquid crystal droplet array that enables simultaneous measurement of these important characteristics of light. Polarization and intensity are measured via division-of-wavefront polarimetry, exploiting the droplets' spatially varying birefringence, while the phase is reconstructed by treating each droplet as a separate microlens in a Shack-Hartmann-like wavefront sensor configuration. To demonstrate the system's performance, we characterize aberrated dual-wavelength beams carrying distinct intensity, phase, and polarization information, confirming accurate retrieval of the optical field profiles for both spectral components.**

## Introduction

The amplitude, phase, and polarization of light are fundamental parameters that govern the propagation and structure of optical fields. Accurate measurement of these parameters is essential for advancing photonic technologies and for understanding complex optical phenomena, with applications ranging from metrology[1,2], microscopy[3-5] and astronomy[6-8] to optical communication[9], remote sensing[10] and biomedical diagnostics[11,12]. Traditional optical techniques can measure these properties either separately or simultaneously using complex devices that are challenging to construct or fabricate[13-30]. A cost-effective approach that features faster and simpler fabrication of a compact optical device would offer significant advantages for the above applications[31-36].

In response to these evolving requirements, we introduce a compact, and low-cost liquid-crystal (LC) droplet array that enables single-shot measurement of intensity, phase, and polarization. The device is fabricated by inkjet printing, a process that avoids lithography and complex alignment while still offering precise control over droplet size and spacing, making it straightforward to produce uniform arrays in a scalable manner (see details in Method). The polarization sensing principle is based on division-of-wavefront polarimetry (DoWP), where different regions of the incident beam encode distinct optical information through interaction with each droplet's spatially varying birefringence profile, enabling simultaneous retrieval of all vectorial components of the optical field.

In parallel, phase information is obtained by treating each droplet as a microlens in a Shack-Hartmann wavefront sensor (SHWS) configuration[37,38]. Its curved interface introduces spatially varying optical pathlengths that focus and deflect the transmitted light, generating focal spot displacements that are analysed across the array to reconstruct the complete phase distribution of the incident wavefront. Combined with an RGB imaging detector, the platform simultaneously captures intensity, polarization, and phase information and can naturally extend to multi-wavelength sensing within a single exposure.

In this paper, we validate the proposed sensing concept using a dual-wavelength optical system in which each wavelength carries independently modulated intensity, phase, and polarization information. The RGB detector separates the two wavelengths and retrieves their spatial field distributions within a single camera exposure, confirming simultaneous recovery of all optical parameters in a sufficiently high quality. This proof-of-concept demonstration verifies the feasibility of the LC droplet array for real-time, full-field optical characterization and establishes a practical foundation for future development in adaptive optics, optical communication, and imaging systems.

## Results

### 1.1 Concept

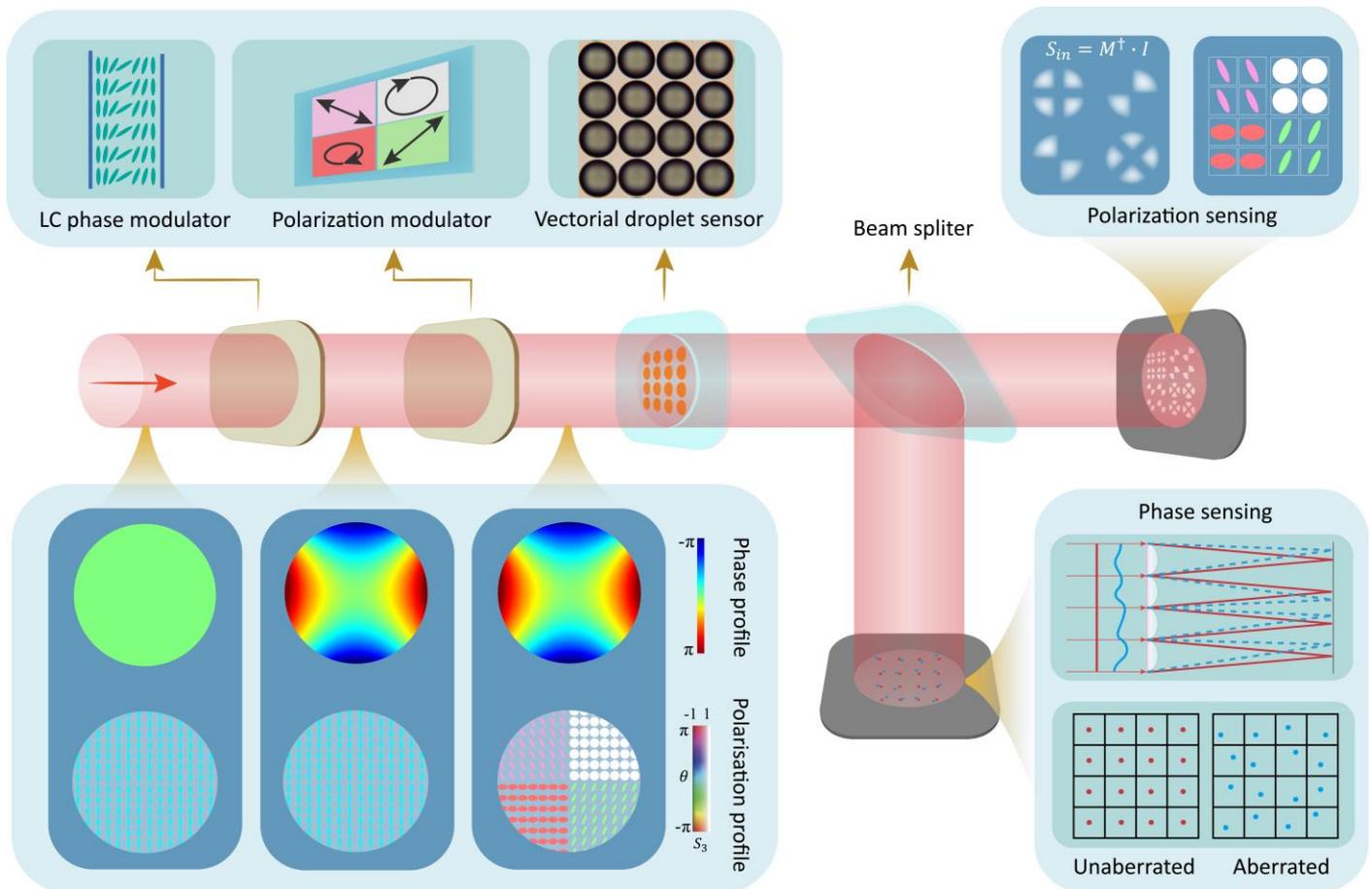

**Fig. 1 Schematic overview of the liquid crystal (LC) droplet array-based sensing system for simultaneous intensity, phase, and polarization measurement of monochromatic light.** The incident beam is sequentially modulated in phase-using a spatial light modulator (SLM) or an LC wedge cell and in polarization via a four-region division waveplate that generates spatially varying polarization states, with the corresponding phase and polarization profiles presented at different stages of the incident beam before the printed LC droplet array. The modulated beam is then directed onto the printed LC droplet array. The configuration for the incident light is identical for both single- and dual-wavelength experiments presented in this work. Throughout this paper, Stokes fields are depicted using hue to signify the azimuthal angle $\tan\theta = S_2/S_1$ and saturation to represent height $S_3$. A non-polarizing beam splitter separates the transmitted light into two optical paths: one for phase sensing, where focal spot displacements are analysed using the Shack–Hartmann principle, and the other for polarization sensing, where the internal intensity distribution within each droplet is recorded and used to retrieve the local Stokes parameters. Together, these channels enable real-time, single-shot reconstruction of the vectorial optical field, including phase and polarization components, across the entire beam profile.

To validate the simultaneous sensing of full vectorial field with our inkjet-printed nematic LC droplet array, we designed an experiment (**Fig. 1**) in which a laser beam undergoes sequential phase modulation (through an LC phase modulator) and polarization modulation (by passing through a segmented waveplate) prior to interacting with our printed LC droplet sensor. Upon transmission through the droplet array, the modulated beam is then divided into two optical paths, each path being responsible for a different sensing function. Firstly, for the intensity and polarization path, the spatial intensity patterns across the droplet array are recorded, from which the total intensity can then be obtained directly and the local Stokes vectors are retrieved using a pre-calibrated system matrix (see details in **Supplementary Note 1**). For the second path, denoted the phase path, the LC droplet array generates an array of focal spots, where displacements of the spots are subsequently analysed to extract local phase gradients so as to reconstruct the wavefront using Zernike polynomial fitting (see details in **Supplementary Note 2**).

By processing the two detection channels simultaneously, the system provides real-time access to the beam's intensity, phase, and polarization. In our design, two different wavelength light sources are used simultaneously with independent phase and polarization modulation in each channel. The RGB detector separates the spectral components, allowing independent reconstruction of all field parameters within a single exposure. Note that in a later section we focus on the polarization and phase retrieval procedures, with the intensity obtained through brightest-point analysis, as detailed in Supplementary Note 2.

## 2.1 Polarization measurement

The polarization measurement procedure involved using a dual-wavelength configuration consisting of red ($\lambda = 633$ nm) and green ($\lambda = 532$ nm) laser sources, each undergoing independent polarization modulation. Polarization modulation was achieved with a custom-designed four-division waveplate (FWP) **(Fig. 2a)**, assembled by bonding half-wave plates (HWP) and quarter-wave plates (QWP) of different orientations into four spatially-equal quadrants, each generating a distinct and well-defined polarization state. After passing through the FWP, the two beams, each carrying four distinct polarization states, were combined into a single co-propagating path and directed onto the LC droplet array sensor. Transmission through the droplet array resulted in the generation of wavelength-dependent intensity patterns, which were then recorded by an RGB camera that separated the red (633 nm) and green (532 nm) channels.

In each channel, interaction between the incident state of polarization (SoP) and the droplets' spatially varying birefringence results in a characteristic distribution of the intensity across the droplets. These intensity distributions were analysed independently for the red and green channels (See details in Supplementary Note 1). The local intensity values within each droplet image were then processed through a pre-calibrated Mueller matrix inversion to retrieve the spatially resolved Stokes parameters, thereby reconstructing the full SoP for each spatial region of the beam. As can be seen in Fig. 2a, the retrieved polarization ellipses from the four regions in both channels closely matched the expected input states, confirming the reliable performance of the droplet array's polarization measurement.

A quantitative comparison is presented in Fig. 2b, where the three normalized Stokes parameters ($S_1, S_2, S_3$) for each quadrant are plotted, using the red channel results as a representative example. Red diamonds indicate the mean experimental values (including variations among droplets), and the overlaid bars show the reference values extracted from the known FWP states. Across all quadrants and all Stokes components, the two datasets are found to agree within ±3%, thereby validating the reliability of the LC droplet array polarimeter for real-time, spatially resolved polarization measurement for a single wavelength light source.

In addition to polarization information, the total intensity was obtained directly from the same CCD images used for Stokes reconstruction. The mean transmitted intensity within each droplet region was extracted to form the corresponding element of the intensity vector, and a reference frame recorded without the analyzer was used for normalization to correct illumination nonuniformity (see Detail in Supplementary Note 1). This integrated process enabled simultaneous acquisition of both the total intensity and polarization data within a single fixed exposure time.

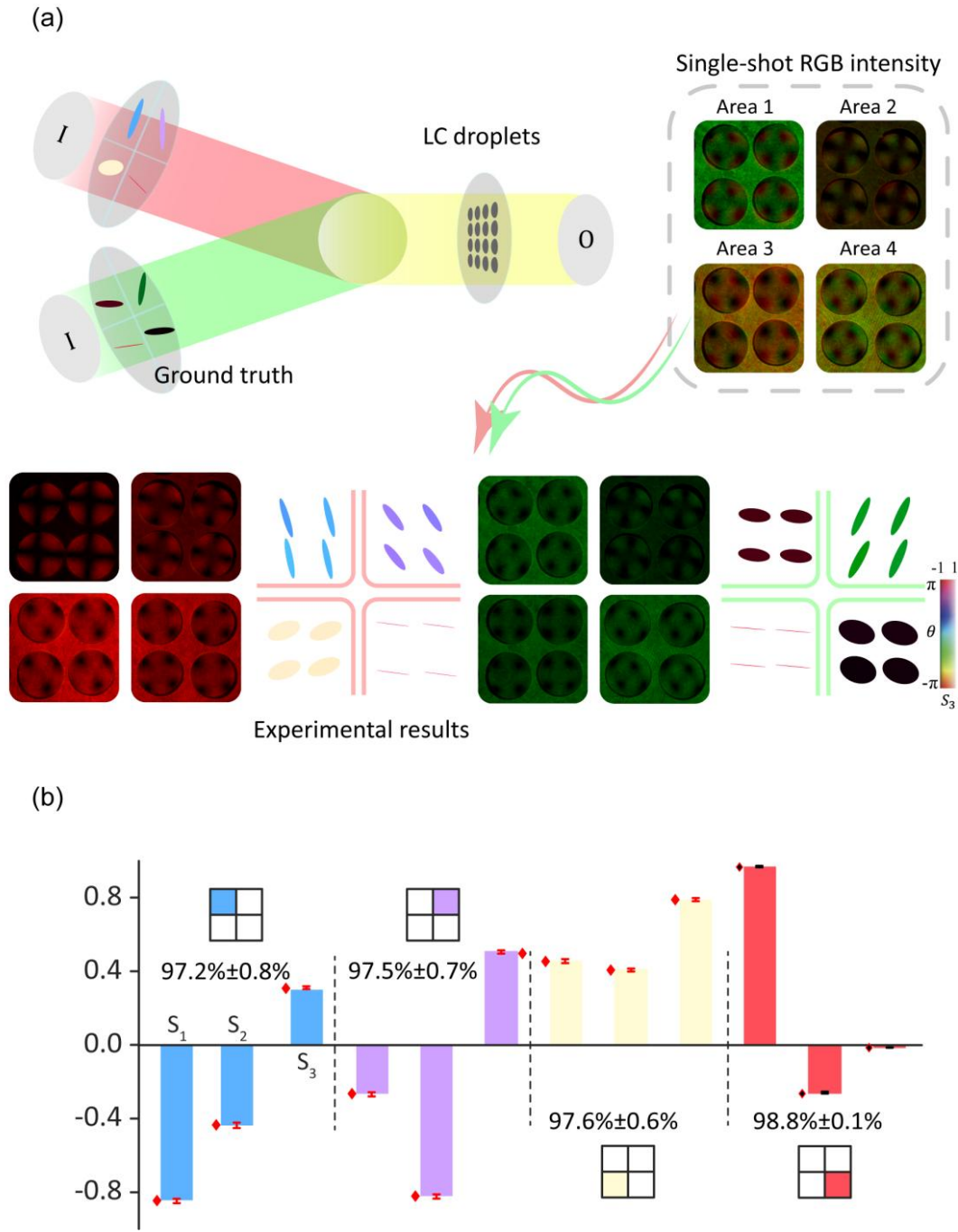

**Fig. 2 Dual-wavelength polarization measurements using the LC droplet array sensor.** (a) The diagram illustrates the measurement configuration, the different input polarization states, and representative images captured on the CCD camera. The incident beam contains red (top beam) and green (bottom beam) light, each one is spatially divided into four quadrants with four distinct polarization states. The output was then recorded by an RGB CCD camera, and the captured image was separated into (i) red (left side of the figure) and (ii) green (right side of the figure) channels for independent analysis (I: Input, O: Output). Shown alongside four representative images captured by the CCD camera are examples images of the droplets corresponding to each of the four quadrants and the corresponding reconstructed polarization ellipses after decomposition into the red and green channels. The results demonstrate the system's ability to resolve different states of polarization (SoP) for both wavelengths. The colour bar represents the Stokes field, consistent with Fig. 1. (b) Quantitative comparison between the droplet-array measurements (red diamonds, averaged over four droplets per polarization quadrant; error bars denote the spread of values over the four droplets) and the reference values derived from the known FWP states (bar charts). The percentages indicate the overlap between the mean experimental results and the corresponding reference values, confirming the consistency of all Stokes parameters across the four input SoPs.

## 2.2 Phase measurement

Using our system, measurement of the phase was carried out as follows. To begin with, a reference image without phase modulation was recorded at the focal plane to establish the baseline positions of the focal spots. Next, a dual-wavelength configuration was employed, with a $\lambda = 633$ nm beam phase-modulated by a spatial light modulator (SLM) to impose defined Zernike aberrations, and a $\lambda = 532$ nm beam modulated by a nematic LC wedge cell, which introduced a controlled tilt through its adjustable orientation. Three phase states – vertical tilt, horizontal tilt, and their combination – were applied to the green beam, while the red beam was sequentially assigned aberrations in the form of defocus, oblique astigmatism, and vertical astigmatism. Subsequently, the two beams were combined into a single optical path and directed onto the LC droplet array. Transmission through the array generated two sets of focal spot patterns, captured by the RGB camera, which separated the red and green channels for independent analysis. The focal spots corresponding to the two wavelengths appeared at different positions within each droplet due to their distinct aberration profiles, as illustrated schematically in **Fig. 3(a)**. Lastly, the spot displacements relative to the reference image were calculated and used to reconstruct the phase distribution across the pupil (See details in Supplementary Note 2).

Figure 3(b) shows the reconstructed 2D phase profiles at three different phase states ($T_1$, $T_2$, $T_3$), with the red and green channels plotted separately. The recovered phase maps reproduce the imposed aberrations in both spatial distribution and relative magnitude. To further assess the reconstruction accuracy beyond these general results, the red channel was selected as a representative case. Four additional Zernike aberrations were sequentially applied to the 633 nm beam using the SLM and the results are summarized in Fig. 3(c), where the reconstructed coefficients correspond to the applied modes, and the inset images compare the reconstructed and theoretical phase maps, showing consistent spatial distribution. These observations confirm that the droplet-array-based sensor can resolve phase aberrations in vectorial beams.

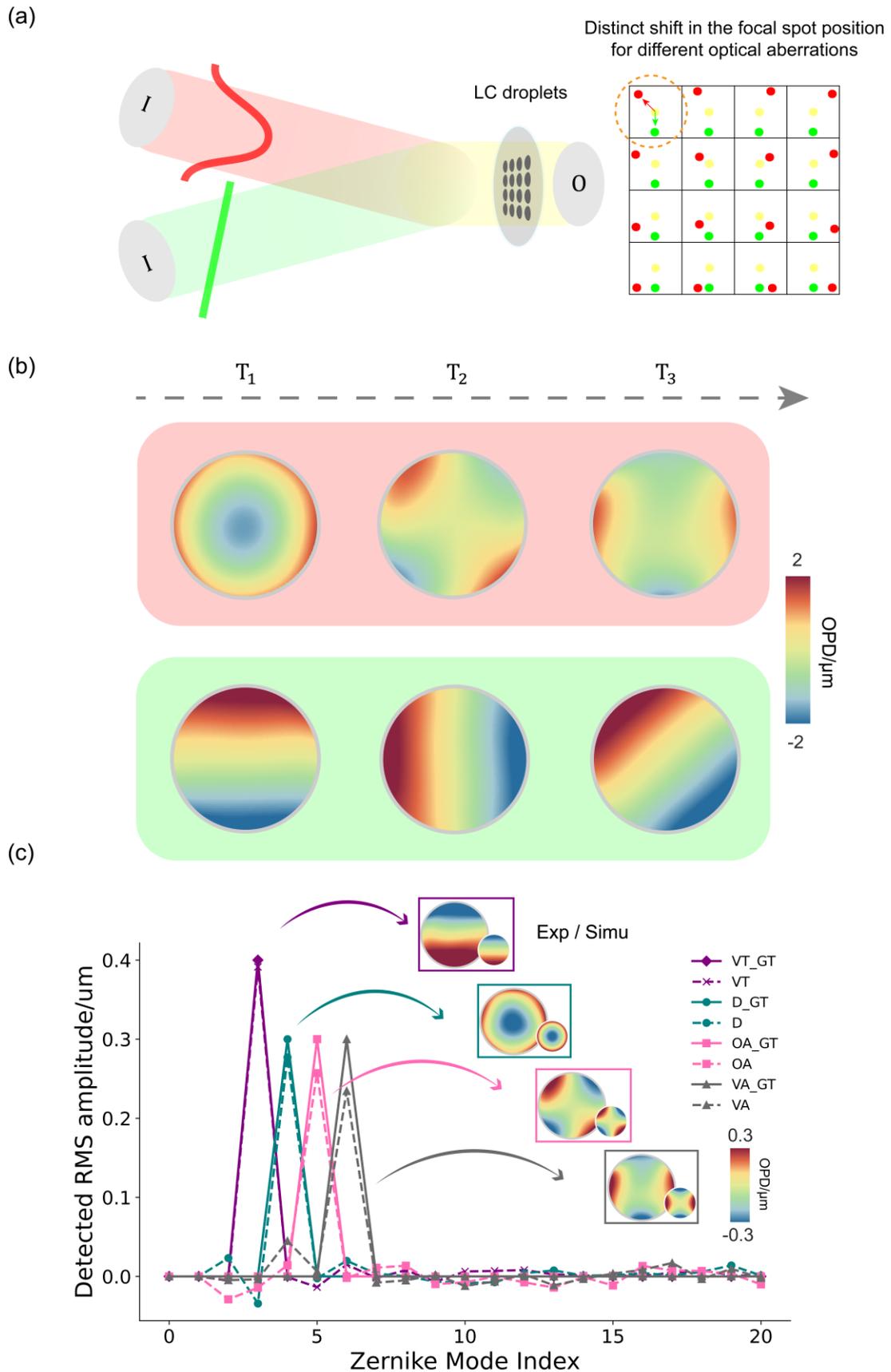

**Fig. 3 Validation of phase sensing performance using the LC droplet array.** (a) Schematic of the dual-wavelength phase sensing measurement process: defocus aberration was applied to the red beam using an SLM, while a fixed vertical tilt was induced in the green beam via a nematic wedge cell. After passing through the droplet array, the focal spot pattern showed distinct spatial displacement for each wavelength (I: Input, O: Output). (b) Optical path difference maps for both red and green channels for different phase states. Each column represents a different phase state ($T_1 - T_3$), demonstrating the system's capability for simultaneous, multi-wavelength wavefront sensing. (c) Reconstructed Zernike coefficients for four additional aberrations introduced in the red channel using the SLM: vertical tilt (purple; VT),

defocus (green; D), oblique astigmatism (pink; OA), and vertical astigmatism (grey; VA). Solid lines represent experimental results whereas dashed lines denote ground truth values obtained from the corresponding simulated Zernike phase aberrations. Insets show simulated (small circle) and measured (large circle) wavefront profiles. The Root Mean Square Error (RMSE)[39], expressed as a percentage of the corresponding ground truth coefficient, was found to be 1.6% (VT), 4.1% (D), 4.8% (OA), and 6.2% (VA).

**Discussion and Conclusions**

The LC droplet array methodology presented in this work provides a compact, easy-to-fabricate, and cost-effective approach for full optical field measurements. By integrating DoWP polarization sensing principle and a SHWS into a single platform, the device simultaneously captures polarization, phase, and intensity information, and can be readily extended to multi-wavelength operation using an RGB camera. It should be emphasised that our demonstrations of the sensing procedure are proof-of-concept, and that a detailed error analysis together with targeted optimization will be necessary before the technique can be deployed in wider applications.

Further exploration of the system's limitations and optimization strategies are warranted: in polarization sensing, the main sources of error include systematic effects arising from imperfections in the LC droplets, other optical components, and the detector, as well as variations in the incident beam profile. In addition, noise, such as Gaussian or Poisson noise, further degrades the measurement performance. This leads to three potential directions in the future to further enhance measurement accuracy and precision. Firstly, a rigorous calibration pipeline should be implemented in which the input Stokes vectors span the widest possible region of the Poincaré sphere, ensuring accurate estimation of the instrument matrix [40]. Secondly, the droplet geometry can be further refined to take advantage of retardance regions that yield optimal measurement sensitivity; for example, the 132° retardance condition provides a near-optimal condition number and should be preferentially exploited in future designs [14]. Thirdly, the impact of random noise can be reduced by averaging over multiple frames. Beyond conventional averaging, the calibration, optimization, and denoising processes could potentially be combined into a single step using machine-learning models trained directly on imaging data[41].

For phase sensing, it is important to point out that the a key source of error may be the geometric phase introduced by the spatially varying birefringence[42] of the printed LC droplets. Such effects should be considered if further enhance measurement accuracy and precision is to be achieved. Potential strategies could involve inserting a circular-polarization beam splitter after the droplet to isolate the two orthogonal polarization components and to then measure their phase profile separately, thereby enabling full phase reconstruction. Alternatively, it might be possible to pre-characterize the droplet's geometric phase (e.g., rapid in-situ Mueller matrix polarimetry) and use the output from the pre-characterization as a vital reconstruction information for Shack-Hartmann analysis. Separately, future work could involve optimizing the droplet geometry to minimise unwanted geometric phase while retaining the birefringence required for polarization sensing. Moreover, data-driven approaches could be applied, in which neural networks are trained directly on imaging data to infer the true input phase profile while compensating for geometric-phase artefacts.

Beyond the error sources discussed above, there are two strands of further development that could be explored that could result in enhanced performance and extend the functionality in the future. The nematic LC (E7) droplets used here must be at least 330 μm in diameter to ensure sufficient retardance ($0° - 180°$). Such a large droplet size limits the number of droplets that could be arranged within the imaging field, constraining spatial resolution in polarization sensing and sampling density in phase reconstruction. As a result, the polarization maps exhibit coarse spatial averaging, and only low-order Zernike modes could be retrieved in the phase analysis. To overcome these limitations, one potential solution would be to employ nematic LC droplets with a negative dielectric anisotropy, where the application of an electric field compresses the retardance range, thereby facilitating smaller droplet sizes. Increasing droplet density would benefit vectorial sensing by enabling finer spatial mapping of local SoPs and by providing more focal spots for wavefront sampling, thus promoting the retrieval of higher-order Zernike modes. Additionally, extending the technique to incorporate wavelength measurement functionalities would create a full-field sensor for polarization, phase, wavelength, and intensity using a monochromatic camera. This would rely on calibrating the wavelength-dependent spot profile at a fixed detection plane, potentially enhanced by machine-learning models. Such advancements would unlock new opportunities in adaptive optics, hyperspectral imaging, structured light metrology, and biomedical diagnostics, potentially offering deeper insights into optical phenomena and facilitating more versatile, high-resolution analysis.

Overall, we have demonstrated a printed LC droplet array-based vectorial field sensor that combines compact design, low cost, and ease of fabrication for full-field optical sensing. With further refinement, this approach holds strong potential as a versatile tool for both scientific and technological applications.

**Methods**

The LC droplet array was fabricated using a drop-on-demand inkjet printing method, chosen for its precision and scalability in producing well-defined arrays. The process involved spin coating glass substrates with a lecithin solution (0.02% lecithin and 99.98% Isopropyl Alcohol) to form a homeotropic alignment layer, which ensures that the LC director aligns perpendicularly to the glass substrate. Inkjet printing was performed using a MicroFab Jetlab II system, equipped with an 80 µm-diameter nozzle. The printhead was operated at 70°C, a temperature intentionally set above the nematic to isotropic liquid transition temperature ($T_c$) of the nematic mixture (E7, Synthon Chemicals Ltd; $T_c$ = 58°C). This elevated temperature mitigates viscosity anisotropy, thereby enhancing the uniformity and stability of the printed droplets. LC droplets were then precisely deposited onto the prepared glass substrate in a 2D array, as illustrated in **Fig. 4**. The droplet formation and deposition were governed by a trapezoidal voltage waveform, enabling precise control over the process. The droplets were accurately positioned using a computer-controlled XY motorized stage, maintaining a uniform spacing of 10 µm between droplets. The droplet size could be adjusted either by changing the nozzle or by depositing multiple droplets at the same location, allowing for precise customization based on specific experimental needs, in this case 20 droplets were deposited at each site to produce final droplets with a diameter of approximately 330 µm (Fig. 4(d)).

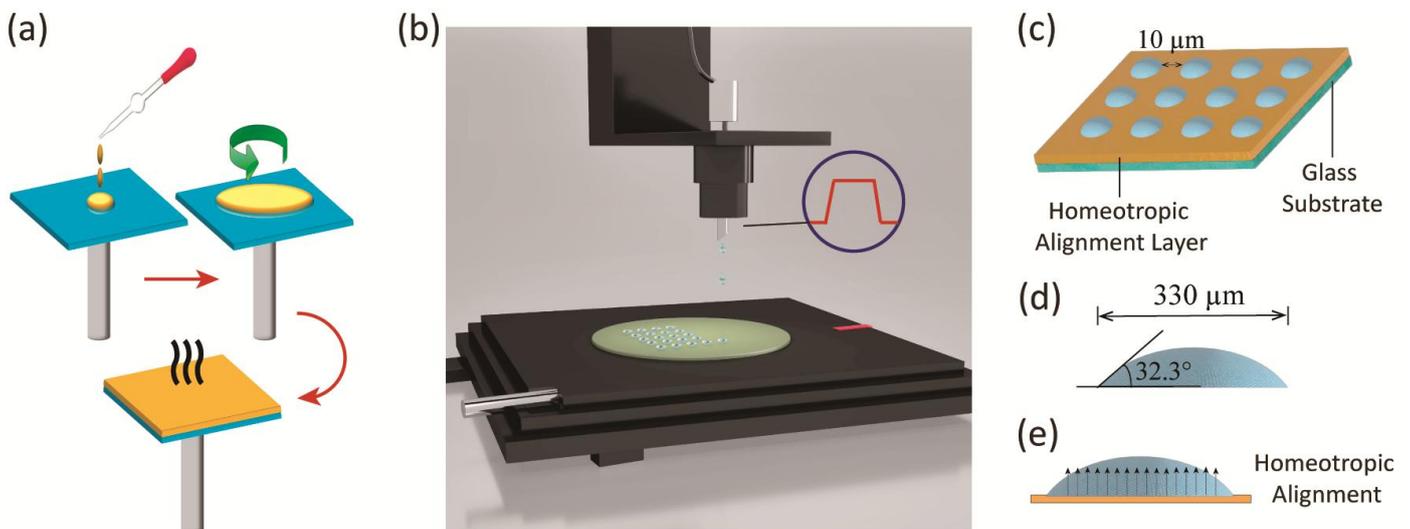

**Fig. 4 Schematic and structural layout of the nematic LC droplet array fabrication process.** (a) Illustration of the process of fabricating homeotropic alignment layer. (b) Illustration of the MicroFab inkjet printing system used to deposit nematic LC droplets in an array arrangement. (c) Schematic of the printed sample, showing nematic LC droplets arranged in an array with 10 $\mu m$ spacing, positioned on a homeotropic alignment layer supported by a transparent glass substrate. (d) Cross-sectional view of a single nematic LC droplet, indicating a contact angle of 32.3° and a diameter of 330$\mu m$. (e) showing the effect of the homeotropic alignment layer on the LC director orientation, which enforces perpendicular alignment at the droplet-substrate interface.

**Acknowledgments**

AACP acknowledges funding from the NSF/CBET-EPSRC (Grant Nos. EP/W016036/1 and EP/S029966/1). and the John Fell Fund via a Pump-Priming grant (0005176). C.H. acknowledges support from St John's College, the University of Oxford, and The Royal Society (URF/R1/241734). WK, SJE and SMM acknowledge the EPSRC-UK for a research grant (EP/W022567/1).

**Competing interests**

The authors declare no competing interests.
**Additional Information**

Correspondence and request for materials should be addressed to C.H. and S.M.M.